\documentclass{article}

\usepackage[dblblindworkshop, final]{ia}
\usepackage{parskip}
\usepackage{listings}
\usepackage{wrapfig}

\usepackage[utf8]{inputenc} % allow utf-8 input
\usepackage[T1]{fontenc}    % use 8-bit T1 fonts
\usepackage{hyperref}       % hyperlinks
\usepackage{url}            % simple URL typesetting
\usepackage{booktabs}       % professional-quality tables
\usepackage{amsfonts}       % blackboard math symbols
\usepackage{nicefrac}       % compact symbols for 1/2, etc.
\usepackage{microtype}      % microtypography
\usepackage{xcolor}
\usepackage{graphicx}
\usepackage{amsmath}         % colors

\workshoptitle{Efficient Reasoning}

\title{iOS as Acceleration}

\author{
 Alexander K. Chen \\
 Independent High School Researcher \\
 \texttt{alexander.kai.chen@gmail.com} \\
}

\begin{document}

\maketitle

\begin{abstract}
Practical utilization of large-scale machine learning requires a powerful compute setup, a necessity which poses a significant barrier to engagement with such artificial intelligence in more restricted system environments.
While cloud computing offers a solution to weaker local environments, certain situations like training involving private or sensitive data, physical  environments not available through the cloud, or higher anticipated usage costs, necessitate computing locally.
We explore the potential to improve weaker local compute systems at zero additional cost by taking advantage of ubiquitous yet underutilized resources: mobile phones.
Specifically, recent iOS phones are equipped with surprisingly powerful processors, but they also face limitations like memory constraints, thermal throttling, and OS sandboxing.
We present a proof-of-concept system demonstrating a novel approach to harness an iOS device via distributed pipeline parallelism, achieving significant benefits in a lesser compute environment by accelerating modest model training, batch inference, and agentic LRM tool-usage.
We discuss practical use-cases, limitations, and directions for future work.
The findings of this paper highlight the potential for the improving commonplace mobile devices to provide greater contributions to machine learning.
\end{abstract}

\section{Introduction}

Running practical large-scale model training and inference requires devices with powerful compute.
Such devices are at times prohibitively expensive, leaving smaller research labs, universities, and individuals limited to weaker compute setups.
Cloud computing alternatives are not always readily available, e.g.\ in situations where outsourcing compute to third parties risks data privacy or is physically infeasible, like due to required local hardware interactions.

However, thousands of consumer mobile devices are shipped each year with increasingly powerful processors, now capable of running substantial local machine learning tasks independently~\cite{xiao2025understandinglargelanguagemodels}; we believe they have the potential to contribute more significantly at another level.
For the sake of this paper, we focus specifically on iOS mobile phones, as newer Apple iPhones feature consistent high-performing processors and optimized frameworks to interact with them.
These devices, while equipped with relatively powerful chips, are limited by memory constraints, thermal throttling, and OS sandboxing, restricting their ability to run training or inference for larger models by themselves~\cite{patil2022poettrainingneuralnetworks}.

In this paper we propose leveraging a pipeline parallel distributed system, letting iOS devices only store small partitions of model weights to still meaningfully contribute their compute, circumventing mobile memory limitations.
This worked well for training and inference with the modestly sized \texttt{ResNet-34}~\cite{DBLP:journals/corr/HeZRS15} with an old 2013 Intel Xeon desktop as integrating an iPhone 16 reaped a ~44\% decrease in training time, detailed in the Results (section~\ref{sec:results}).
However, we found our iOS worker impractical when attempting training on more significant models like LLMs.
This prompted us to pursue relatively lighter computation tasks, leading us to attempt parallelizing tool usage computation with agentic LRMs reasoning.
Further, iOS devices provide a unique suite of physical sensors and on-device features, such as LiDAR 3d-sensing, audio processing, and GPS positioning, which could serve as potential cost-effective paths of exploration towards embodying agentic AI\@.
We experimented with a scenario using \texttt{Qwen3-8B}~\cite{yang2025qwen3technicalreport} on a 2023 Macbook Pro and a mocked vector database search operation and show a theoretical elimination of idle-time waiting for tool results via tool parallelism.
Finally, we discuss limitations and directions for future work in Discussion (section~\ref{sec:discussion}).

\subsection{Device Benchmarks}
\label{subsec:devbench}

Mobile compute power has progressed rapidly over the past years, reaching or surpassing capabilities of lower-end computer hardware.
Table~\ref{tab:bench} compares some iOS integrated chips with other mobile, edge device, and lower-end computer processers.

\begin{table}[h]
    \centering
    \begin{tabular}{lcccc}
        \toprule
        \textbf{Name} & \textbf{Year} & \textbf{TFLOPS} & \textbf{Physical RAM} & \textbf{Neural Engine} \\
        \midrule

        Snapdragon 8 Gen 2 & 2022 & 3.4816 & Varies, 8--16 GB & -- \\
        Jetson AGX Orin 32 GB & 2023 & ~3.33 & 32 GB & -- \\
        Apple M2 (iPad) & 2022/23 & 2.918 & Varies, 8--24 GB & 15.8 TOPS \\
        NVIDIA GeForce GTX 1650 & 2019 & 2.849 & 4 GB & -- \\
        Apple A18 Pro (iPhone 16 Pro) & 2024 & 2.289 & 8 GB & 35 TOPS \\
        \textbf{Apple A18 (iPhone 16)} & 2024 & 1.907 & 8 GB & 35 TOPS \\
        NVIDIA GeForce GTX 1050 & 2016 & 1.862 & 2 GB & -- \\
        Jetson Orin Nano Developer Kit & 2023 & 1.28 & 8 GB & 40--67 INT8 TOPS \\
        Intel Core i5--10400F & 2020 & 0.825 & Varies, 128 GB Max & -- \\
        \textbf{Apple A13 Bionic (iPhone 11 Pro)} & 2019 & 0.63 & 4 GB & 5 TOPS \\
        Snapdragon 6 Gen 1 & 2022 & 0.346 & Varies, 6--8 GB & -- \\
        \textbf{Intel Xeon E3-1225 v3 CPU} & 2013 & 0.061 & Varies, 32 GB max & -- \\

        \bottomrule
    \end{tabular}
    \caption{Assorted devices compute comparison~\cite{appleM2A13, gtx1650, appleA18, intelXeon, qualcomm_snapdragon6gen1_nanoreview, qualcomm_snapdragon8gen2_siliconcat, nvidia_gtx1050_techpowerup, nvidia_jetson_agx_orin32_siliconcat, nvidia_jetson_orin_nano8_siliconcat, intel_corei5_10400f_sysrqmts}.  Bolded devices were used during experimentation.}
    \label{tab:bench}
\end{table}

Note: iOS physical RAM specifications are not indicative of how much RAM a given application can consistently use, due to sandbox restrictions.
For example, an iPhone 11 Pro with 4 GB of physical RAM can force quit an application if it uses more than roughly 2 GB of RAM\@.

\section{Related Work}
As advanced models have continued to grow in size, it has become practically impossible for singular devices to efficiently train or run models due to memory and compute limitations.
Thus, parallelism techniques have been explored like pipeline parallelism~\cite{DBLP:journals/corr/abs-1806-03377, DBLP:journals/corr/abs-1811-06965} and model parallelism~\cite{shoeybi2020megatronlmtrainingmultibillionparameter} which split models across multiple devices to take advantage of combined memory and compute overlapping.

Edge computing also garnered significant interest as distributed training gained more traction and widespread use.
Specifically concerning mobile devices, federated learning has been explored as mobile devices can serve as compute resources for collecting and adaptively processing information to cumulatively train models~\cite{wang2019adaptivefederatedlearningresource, he2024fedkitenablingcrossplatformfederated}.

This paper explores a novel intersection between parallelism and mobile edge computing.
We are at a point where mobile phones are powerful enough such that techniques typically applied to more powerful devices to enhance compute can be applied with such edge devices like mobile phones.
We can apply such techniques to more directly enhance larger-scale model training and inference systems through these devices.

\section{Methods}

\begin{wrapfigure}{r}{0.3\textwidth}
    \centering
    \vspace{-20mm}
    \includegraphics[width=0.3\textwidth]{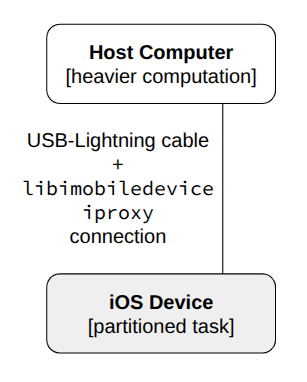}
    \caption{Prototype system.}
    \label{fig:sysgram}
\end{wrapfigure}

\hyphenchar\font=-1

\subsection{System Overview}
To demonstrate our concept, we developed a proof-of-concept application framework, diagrammed in Figure~\ref{fig:sysgram}.
A host computer running its share of compute in PyTorch directs computation offload through a Python interface while wired to an iOS device running our Swift mobile application.
We note that our techniques may not be the most optimal, but are sufficient for demonstrating tangible results.

\subsection{Communication}
\label{subsec:comms}
\begin{figure}[h]
    \centering
    \includegraphics[width=0.7\textwidth]{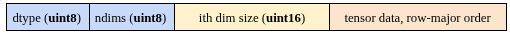}
    \caption{Protocol for communicating tensors between devices.  Datatypes for dimension-related values can be adjusted to accomodate larger tensors.}
    \label{fig:comprot}
\end{figure}

For data communication between the iOS worker and host computer, we use Swift’s Network framework and \texttt{libimobiledevice}\('\)s  \texttt{iproxy}~\cite{libimobiledevice} to communicate bytes through a TCP socket port between wired devices.
We establish a simple communication protocol for sending tensors from device to device by sending datatype, shape information, then raw values sequentially, illustrated in Figure~\ref{fig:comprot}.

\subsection{Mobile Compute}
\label{subsec:mobilecompute}
Our mobile application utilizes Metal Performance Shaders Graph~\cite{mpsgraph}, a subframework of Apple’s hardware-accelerated graphics Metal framework, which supports some common machine learning operations through a static graph construction interface.
We built model layers by hand, translating from PyTorch implementations into MPSGraph and adapting existing operations (see~\ref{subsec:verification}).

\subsection{Verification}
\label{subsec:verification}
Using this communication protocol, we first verified consistency between the PyTorch and MPSGraph natively implemented operations.
We discovered MPSGraph’s \texttt{dropout} operation was implemented using a different scaling formula than expected ($1/r$ instead of $1/(1-r)$), and \texttt{matmul} behaved differently when broadcasting tensors with more than 3 dimensions; both operations we re-implemented accordingly to match PyTorch.
We established the maximum and absolute differences produced by the rest of the operations we planned on using in both frameworks to be within acceptable error range ($1\times 10^{-7} < \text{all maxs}, \text{all averages} < 1\times 10^{-5}$).

\subsection{Pipelining}
\label{subsec:pipelining}
\begin{figure}[h]
    \centering
    \includegraphics[width=0.8\textwidth]{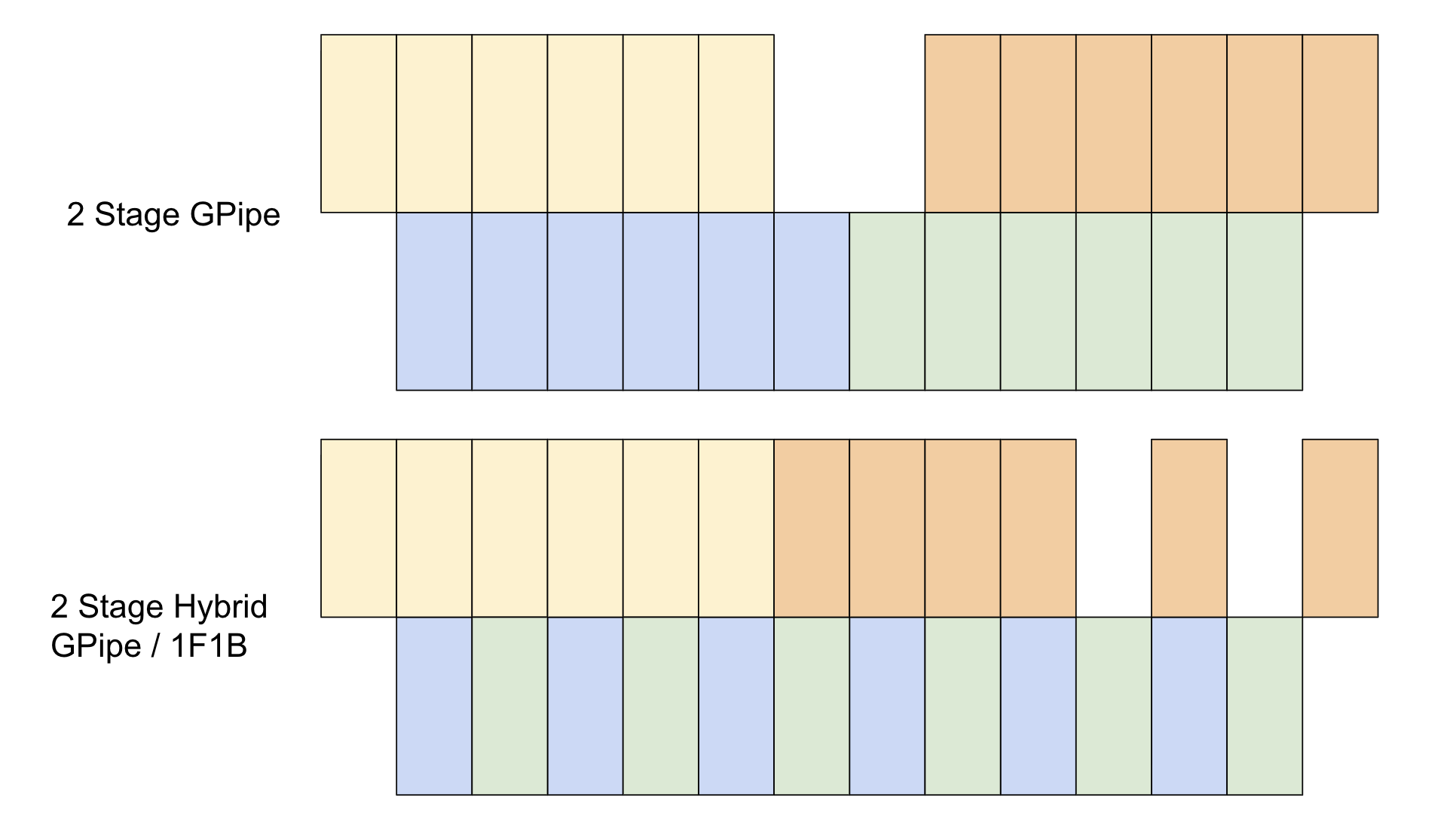}
    \caption{Optimal 2 Stage GPipe vs. Hybrid GPipe/1F1B, time moving to the right. We see the total time taken by both pipelines is equivalent, and the middle bubble in stage 1 of the standard GPipe is spread out at the end of the stage 1 of the Hybrid pipeline.}
    \label{fig:2stagepipe}
\end{figure}

Our first experiment (\ref{subsec:parallel-training-experiment}) attempted to improve model training time via parallelism, and so we implemented a 2-stage pipeline system.
MPSGraph's execution interface is implemented more rigidly than PyTorch: one must explicitly define a static computation graph which will be compiled at runtime, and to which tensor data would be fed through to get results; that is, there are no dynamic compute graphs.
While MPSGraph does support an autograd-like operation, it cannot be run separately from the forward pass and loss calculation operations due to the nature of this static graph construction.
This meant that a standard GPipe~\cite{DBLP:journals/corr/abs-1811-06965} or 1F1B~\cite{DBLP:journals/corr/abs-1806-03377} pipeline (where forward and backward passes are run separately) would not be practically feasible without re-implementing a lot more lower-level operations, so we implemented a hybrid GPipe/1F1B: essentially equivalent to GPipe efficiency-wise for 2 stages, as the bubble is spread out in the backward pass, illustrated by Figure~\ref{fig:2stagepipe}.

\subsection{Tool Usability}
\label{subsec:lrmtool}
Not only are iOS devices equipped with features like advanced positioning / physical and visual sensors as well as built-in intelligence which could serve as interesting tools for LRMs to interact with, their hardware makes them practical workers for potentially expensive tool operations.
We experimented with such in~\ref{subsec:parallel-tool-usage-experiment}
We implemented a simple dot-product vertex database search function which could be interacted with using the same communication protocol as our pipeline system.
Initially, we took 100,000 samples from the AG News dataset~\cite{zhang2015characterlevel} and encoded them using \texttt{all-MiniLM-L6-v2}~\cite{reimers-2019-sentence-bert}, but running this tool took a negligible average of around 10 ms on the iOS device, practically strong but not slow enough to demonstrate our purposes.
Thus, we inflated run time to 5 seconds by using Swift's \texttt{Task.sleep} just to delay returning output from the iOS worker, simulating a long tool usage time for greater visibility during experimentation.
While the host computer could technically have also run the tool operation in parallel during experimentation, our concept shows how in scenarios where less powerful host computers are entirely consumed by resource usages of a model, the iOS worker can still help via compute offloading.

\section{Results}
\label{sec:results}

\subsection{Parallel Training Experiment}
\label{subsec:parallel-training-experiment}

In this experiment, we aimed to reduce the training time of a moderate model on an old consumer-level desktop, specifically a 2013 Intel Xeon desktop (highlighted in~\ref{tab:bench}).
We selected ResNet-34 due to its easily-partitionable 4-layer repetitive-block architecture, and first experimented parallelizing compute onto an iPhone 11 Pro worker, and then secondly an iPhone 16 worker.

Training tests were run on 20 batches of size 128 of ImageNet~\cite{imagenetMini, 5206848} data, with pipeline setups running microbatches of size 16 (8 microbatches per batch).
Running the desktop alone averaged 13104.75 ms per batch, totalling 262.64 seconds across the 20 batches.
We used the implementation from the \texttt{transformers} library~\cite{DBLP:journals/corr/abs-1910-03771} and translated the model layers into \texttt{MPSGraph} operations (see~\ref{subsec:mobilecompute},\ref{subsec:verification}) using FP32.
Recorded timed figures from experimentation are rounded to two decimal places; a table of actual values can be found in Appendix~\ref{subsec:rawvals}.

\subsubsection{iPhone 11 Pro}
We found partitioning right before the 4th residual block of ResNet-34's layer 3 reaped the most benefit in this setup: totaling 203.71 seconds with an average time per batch of 10162.54 ms, this resulted in a ~22\% decrease.
Similarly, this setup could also be used for batch inference.
For training, idle time reached ~63 seconds on device 2 and ~5 seconds on device 1 (from which we can also observe the gap time which would’ve been the buble in the traditional GPipe setup being dispersed in the Hybrid pipeline), totaling ~68 seconds.
This shows that there is still room for improvement, but demonstrates how iOS devices can contribute in training setups.

\begin{figure}[h]
    \label{fig:blocklog}
    \centering
    \includegraphics[width=0.7\textwidth]{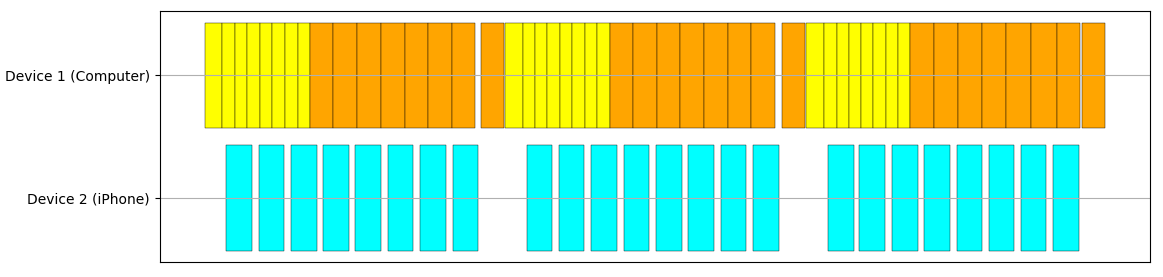}
    \caption{Visual rendering of logs of 3 batch samples from the Layer3Block4 partition training run. Yellow blocks indicate forward passes of a microbatch handled by device 1, orange blocks represent corresponding backward passes, and cyan blocks represent the unified forward+backward pass handled by device 2.  The gaps between cyan block indicate communication overhead.}
\end{figure}

Running inference on 10 batches of size 128 each on the same dataset took a total of 49.78 seconds with an average time per batch of 4399.81 ms for the desktop alone, and only 31.82 seconds with an average time per batch of 2810.50 ms for partitioning ResNet-34 before Layer 3 Residual Block 2, a ~36\% decrease.

\subsubsection{iPhone 16}
We then attempted the same parallel training except replacing the iPhone 11 Pro with an iPhone 16.
This time offloading the entire layer 3 from ResNet, we attained 7308.26 ms average per batch, totalling 146.17 seconds: a 44\% decrease.
This prompted us to also attempt the experiment replacing the Intel desktop with a 2023 MacBook (M2 Max chip), training the same segment of ResNet-34 on CPU-only (no MPS acceleration).
From a baseline of averaging 9008.52 ms per batch and totalling 180.17 seconds, acceleration with the iPhone 16 improved compute time to 6719.06 ms per batch on average with a total of 134.38 seconds: a ~25\% decrease.

We see expected greater significant speedups due to greater memory capacity, raw compute power from the iPhone 16's stronger chip, and also increased communication speed.
Newer iOS devices use the USB-C port, providing a significant speedups to mitigate communication overhead.
Older iOS devices like the iPhone 11 Pro use Lightning cables (USB 2.0, up to 480 Mbps or ~60 MB/s), while newer devices' USB-C ports are USB 3.2 Gen 2, and up to 10 Gbps or ~1.25 GB/s~\cite{appleLightning, appleUSBC}.

\begin{figure}[h]
    \label{fig:multigraph}
    \centering
    \includegraphics[width=0.8\textwidth]{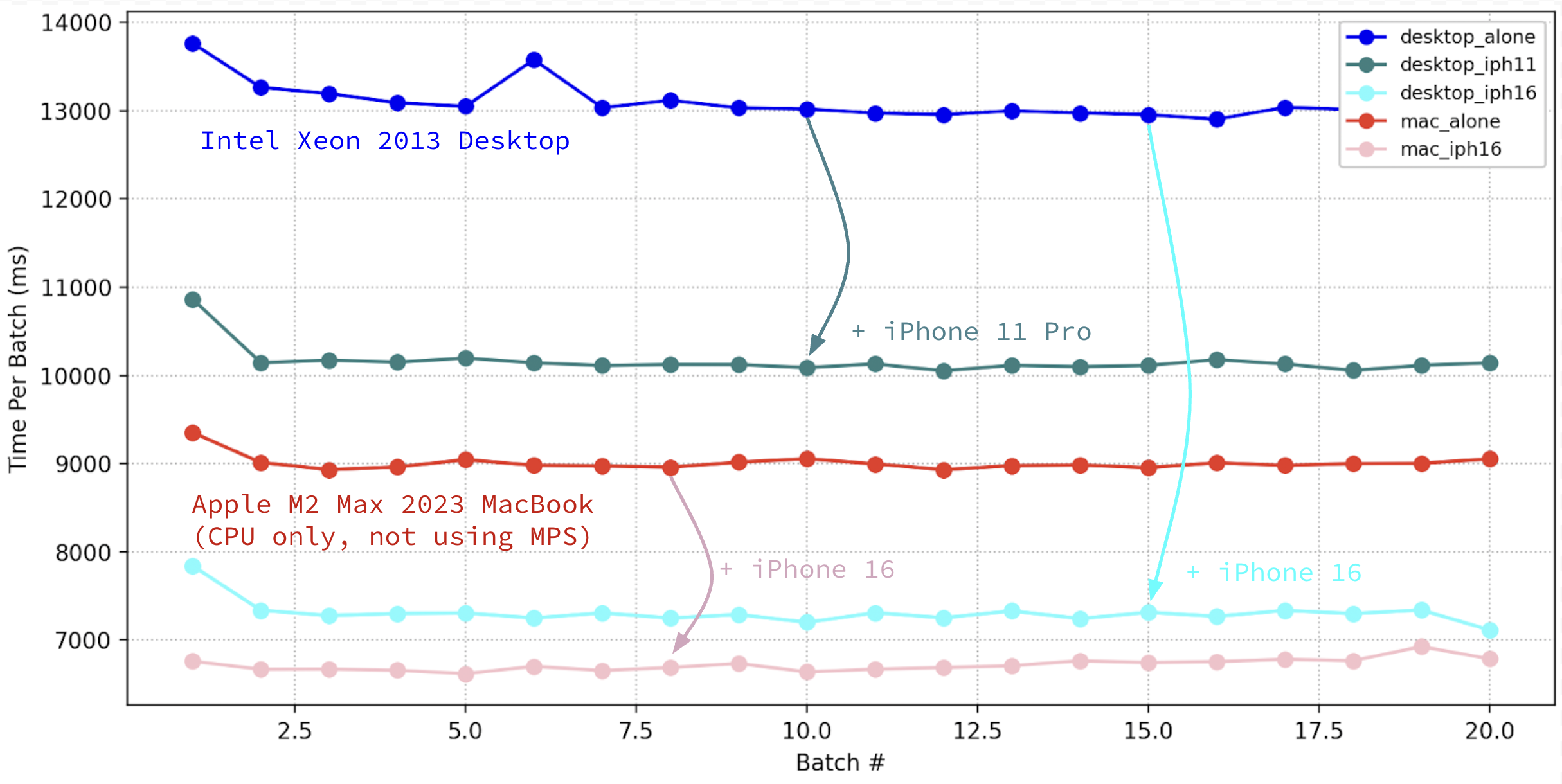}
    \caption{A comparison between all experiments including baseline Intel Desktop, baseline Macintosh CPU-only, and accelerated systems with iPhones}
\end{figure}

\subsection{Thermal Stress Experiment}
\begin{figure}[h]
    \label{fig:energytest}
    \centering
    \includegraphics[width=0.8\textwidth]{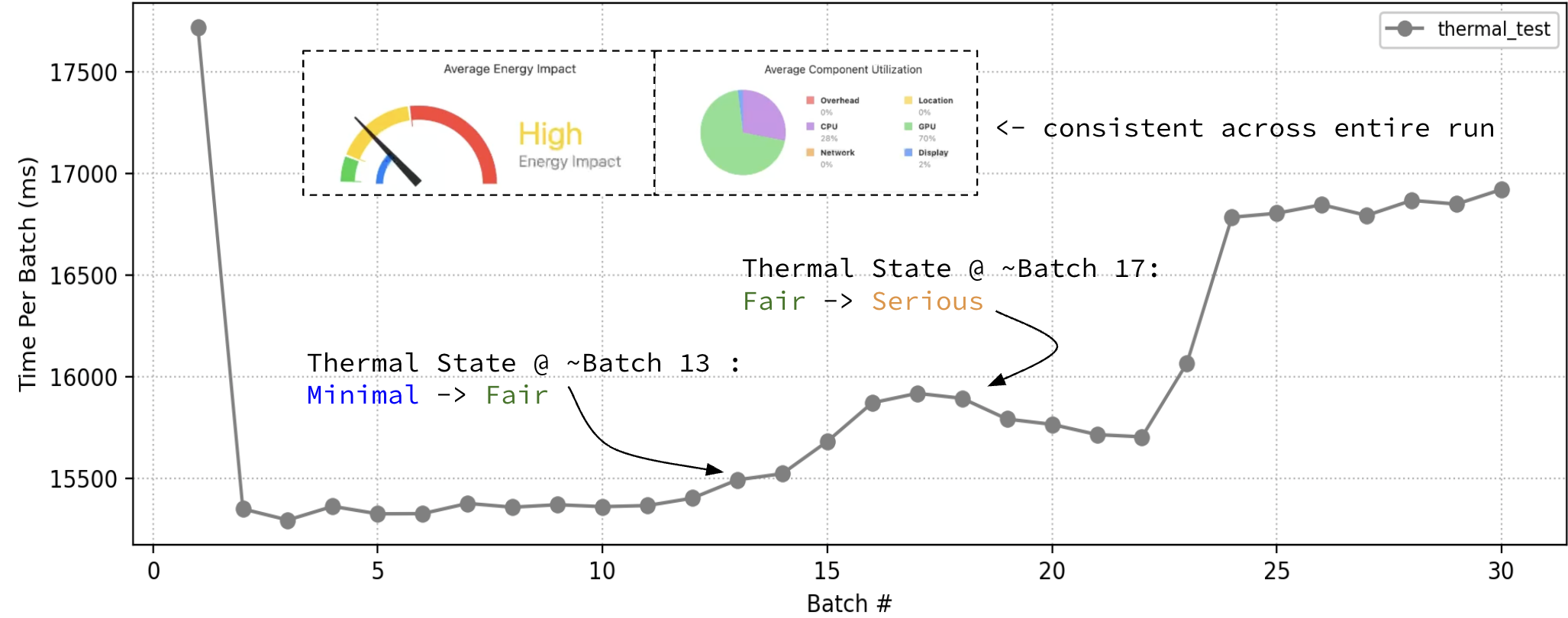}
    \caption{Thermal log, energy test on an iOS, overloading ResNet work; not efficient pipeline, meant to push iOS worker compute}
\end{figure}

Another limitation of iOS devices is their thermal throttling impacts on performance.
In this experiment we explored the impacts of a prolonged strenuous compute task on the iPhone 11 Pro's performance.
From the parallel training experiment, we added the rest of the full layer 3 from ResNet-34 onto the iPhone 11 Pro's load, ensuring its compute load would be fully saturated, running for 30 batches.
Though not an efficient pipeline, the intent of this experiment was to gauge the effects of sustained, heavier computation.
Running this experiment, we saw consistent \texttt{High} energy impact according to the Xcode debugger.
Approaching batch 13, thermal state jumped from \texttt{Minimal} to \texttt{Fair}, and shortly after around batch 17 jumped from \texttt{Fair} to \texttt{Serious}.
Following this, we saw a significant drop in compute speed, of about a couple hundred of milliseconds per batch, indicating thermal throttling had kicked in.
A visual log from the experiment is shown in Figure~\ref{fig:energytest}.

We see effects of sustained computation kicking in, even for training a relatively moderate sized model (only ~162 MB of weights were being trained on the iPhone 11 Pro, taking up a little more than 1.2 GB), displaying the standing issue of thermal limitations.
Discussion about potential solutions is continued in~\ref{subsec:futurework}.

\subsection{Parallel Tool Usage Experiment}
\label{subsec:parallel-tool-usage-experiment}
Tool operations like image generation or large database or internet searches can require larger compute resources to be formed efficiently, more than what might be available on a resource constrained device already running a larger LRM\@.
In the same essence as asynchronous LLM function calling~\cite{gim2024asynchronousllmfunctioncalling}, we can offload tool computation and parallelize it with model generation.
By splitting a tool into two interfaces for the LRM -- one which begins a tool call and one which returns the result (if ready) from the previous tool call -- we can offload the tool operation onto an iOS worker and allow the resource constrained device to continue model reasoning.
To test this idea, we used Qwen Agent framework running Qwen3-8B on a 2023 Macbook Pro, defined the tool protocols using the mock vector database search operation defined in~\ref{subsec:lrmtool}, and ran the agent through the following scenario: a user explicitly queries three search operations and requests a summary for each search's results sequentially.

\begin{figure}[h]
    \centering
    \includegraphics[width=0.7\textwidth]{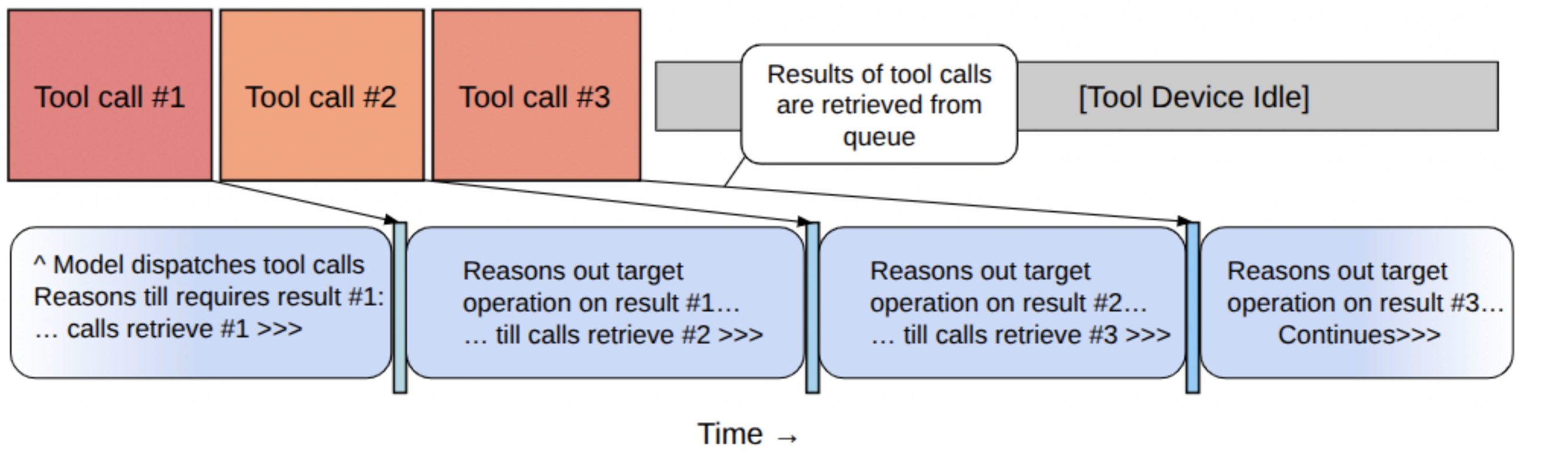}
    \caption{Parallelized tool usage, visualized from time logs.}
    \label{fig:a}
\end{figure}
\begin{figure}[h]
    \centering
    \includegraphics[width=0.7\textwidth]{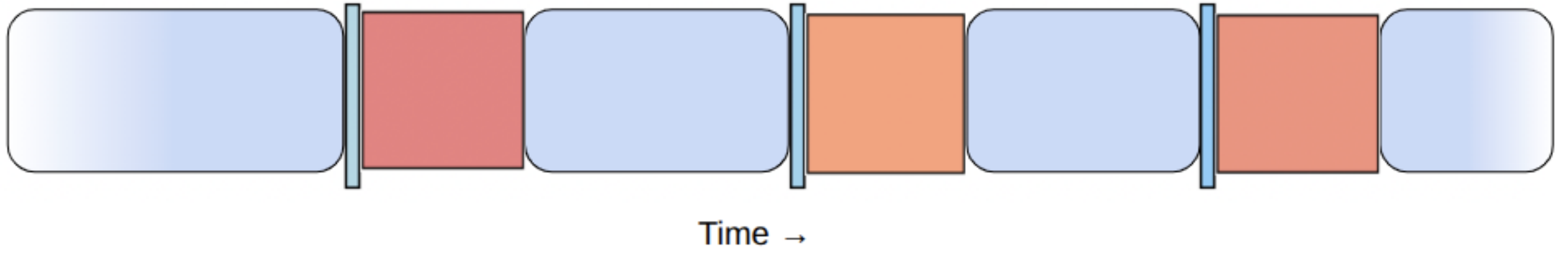}
    \caption{Theoretical non-parallelized tool usage.}
    \label{fig:b}
\end{figure}

Qwen3-8B, with just a given system prompt and detailed tool descriptions, was able to use our split-tool without any further finetuning needed: it followed instructions to interleave search operations, generating summaries from the results of the previous search while having more queued searches run on the iOS worker.
As shown in~\ref{fig:a} and~\ref{fig:b}, offloading tool computation to the iOS theoretically removed it entirely from the overall workflow, excepting the minimal communication overhead of using the interface to call tools.
Full model output, prompts, and tools definitions are provided in~\ref{sec:tech-appen}.

\section{Discussion}
\label{sec:discussion}

\subsection{Limitations}
This paper only highlights the potential to utilize mobile devices for machine learning, providing two example scenarios; it is not an exhaustive record of optimal approaches.
We also only addresses iOS phones specifically; further experimentation involving other mobile devices like Androids or also including tablets should be included for a more expansive survey of consumer edge devices.

To note some limitations of the experiments presented in this paper, the parallel training experiment was performed with a relatively smaller model and only run for twenty batches rather than complete training.
As noted in the thermal experiment, a longer-term experiment may not have reaped as much benefits in aggregate.
The parallel tool experiment also involves a simulated delayed tool call with a smaller, less-powerful model, reducing the potential for more advanced tool calling behavior and reasoning.

We were in part limited in experimentation due to the manual translation of PyTorch operations to MPSGraph.
Translating a large model was incredibly arduous, as model internals and lower-level operations were needed to be reimplemented and verified across interfaces.
While methods for tracing models to generate static compute graphs for model inference in CoreML exist~\cite{apple_coremltools_model_tracing}, these are limited to non-dynamic models and are not available for generating the architecture to train models.

\subsection{Future Work}
\label{subsec:futurework}
We could reasonably expect to see larger RAM allocations and improved GPU / Neural Engines allow for greater acceleration potential; exploring other consumer edge devices including Android phones or tablets, both of which are commonplace and widely available.
While computers in general increase in power, finding methods to utilize already widespread and available devices at minimal costs could reap significant benefits.
Methods to more directly target accelerated hardware could also be explored, as our current method relied on Apple's built-in determination for when to run certain operations on specialized hardware.
The limitations of the static graph also prevented us from running multiple mobile workers to form a pipeline with more than two stages.
Customized frameworks or more flexible implementations could be explored to enable to integration of multiple workers.

The problem of thermal throttling is a significant limitation.
Some potential solutions for consideration could be swapping between multiple iOS workers e.g.\ letting one cool down while another worked almost like pipelining the devices themselves based on thermal usage, or regulating compute loads to short bursts rather than continued work, swapping between only-host and parallelized worker compute.

Finally, these increasingly powerful edge devices may be able to participate in more advanced training environments.
To note some potential use-cases, mobile devices could serve as proxy environments for reinforced learning for game playing or user-recommendation development.
Mobile devices' suite of physical sensors also provide potential value for robotics / embodied AI research, serving as commonly available methods for agents to interact with the world.

\section{Conclusion}
This paper demonstrates how commonplace iOS devices can accelerate model training and inference in modest environments through distributed parallelism.
While facing limitations such as memory constraints and thermal throttling, iOS device chips can still be harnessed to provide additional computation boosts.
This research highlights the potential to utilize ubiquitous devices as powerful compute accelerators and contributing practically to machine learning.

\newpage

\bibliographystyle{plain}
\bibliography{ia}

%%%%%%%%%%%%%%%%%%%%%%%%%%%%%%%%%%%%%%%%%%%%%%%%%%%%%%%%%%%%

\appendix

\section{Appendix}
\label{sec:tech-appen}

\subsection{Parallel Training Times}
\label{subsec:rawvals}

Raw time values (in milliseconds) from parallel training experiments.
All records are batches 1--20, except the thermal test which is batches 1--30.

\begin{lstlisting}[breaklines]
'desktop_alone': [13765.4304, 13264.1586, 13194.2589, 13090.0569, 13049.9169, 13579.1922, 13035.0846, 13118.3392, 13032.2210, 13020.1888, 12973.4548, 12956.3740, 12999.2321, 12975.6014, 12955.8701, 12903.8489, 13038.8358, 13014.0451, 13062.9809, 13065.8304],

'desktop_iph11': [10865.1685, 10144.7933, 10173.3036, 10151.0260, 10195.9800, 10143.4871, 10111.4533, 10123.0546, 10122.1774, 10089.0243, 10129.9788, 10052.4917, 10114.6253, 10099.8297, 10112.9924, 10179.2488, 10130.0227, 10056.3474, 10114.1994, 10141.9436],

'desktop_iph16': [7842.7055, 7337.4474, 7277.5887, 7300.4473, 7306.2833, 7249.9061, 7307.1341, 7249.0506, 7288.8679, 7200.1275, 7309.8252, 7251.9770, 7330.0176, 7243.1087, 7313.9044, 7268.3287, 7334.9983, 7299.6751, 7339.7219, 7114.0900],

'mac_alone': [9352.8128, 9012.3925, 8931.7847, 8962.2284, 9043.8475, 8980.8868, 8972.5937, 8959.1440, 9015.4317, 9054.6023, 8995.7078, 8931.3330, 8976.2855, 8983.7624, 8953.3640, 9009.3956, 8979.2352, 9000.4463, 9002.7686, 9052.3757],

'mac_iph16': [6759.6919, 6668.1087, 6670.1243, 6656.6105, 6618.3534, 6701.9173, 6653.6384, 6688.6338, 6734.3120, 6638.3071, 6669.2123, 6688.2745, 6708.3030, 6765.2090, 6744.3740, 6755.8524, 6781.5692, 6766.0386, 6925.3969, 6787.3247],

'thermal_test': [17720.7760, 15349.7591, 15294.8820, 15362.3798, 15325.4538, 15326.4324, 15376.8889, 15358.1799, 15370.3549, 15360.8573, 15366.2495, 15402.6989, 15492.7669, 15523.2010, 15681.9552, 15871.9805, 15918.7923, 15894.1048, 15792.0616, 15765.8890, 15715.5912, 15704.5098, 16067.0392, 16785.7077, 16805.3755, 16847.6350, 16794.7388, 16868.7144, 16850.5178, 16922.7285],
\end{lstlisting}

\subsection{LRM System Prompt}
The following is the full system prompt passed into the \texttt{Qwen3-8B} model to use our provided tools in the agentic LRM experiment:

\texttt{You are an agent with access to two tools: begin\_search and retrieve\_search.
Begin search starts a vector db search with the given query for topk documents,
retrieve search returns the result of the oldest not-yet-retrieved vector db search (FIFO queue style).
You can do other tasks like summarization in between tool calls, for example:
1. call begin search for query 1
2. call begin search for query 2
*now both are runnning
3. call retrieve\_search for query 1
4. instead of doing another tool call, just summarize the results from retrieve\_search
5. call retrieve\_search for query 2
See how you can do summarization (which is not a tool) in between tool calls? you do not have to run all tool calls together.
But still remember to run all tool calls which the user specifies.
}

\subsection{Tool Prompts}
The following are the tool prompts used by building upon Qwen Agent's \texttt{BaseTool}:

\textbf{name:}

\texttt{vector\_db\_begin\_search}

description:

\texttt{Begins a vector db search through documents from the 2015 AG News dataset to produce `k` most-similar documents.
Search results can be retrieved by using the `vector\_db\_retrieve\_search\_result`.
This tool enables you to continue on other tasks (like summarization) while searches are running.
}

parameters:

\texttt{\{'name': 'query', 'type': 'string', 'description': 'The search query for vector db search.', 'required': True\}}

\texttt{\{ 'name': 'k', 'type': 'int that fits in uint8', 'description': 'The number of documents to return', 'required': True\}}

\textbf{name:}

\texttt{vector\_db\_retrieve\_search\_result}

description:

\texttt{Returns the result of the oldest not-yet-retrieved vector db search tool call, returning the `k` most-similar documents. Functions like a FIFO queue.}

\subsection{User Prompt}
The user prompt sent to the \texttt{Qwen3-8B} model during the agentic LRM experiment.

\texttt{Follow these instructions exactly, one after another sequentially:
1. use `begin\_search` for information about Google\'s search engine
2. use `begin\_search` for information about Apple\'s iPod
3. use `begin\_search` for information about Microsoft\'s Windows
4. use `retrieve\_result` once to get the result of the Google query
. summarize the result of the Google query.
6. after that, use `retrieve\_result` once to get Apple query result
7. summarize the result of the Apple query.
8. after that summary, use `retrieve\_result` one more time
9. summarize the result of the Microsoft query.
It is ok for you as an assistant to run a tool, use its result in your basic reasoning, then run another tool call - you don\'t have to run all at once.
Once all of this is done, just return the three summaries you had written.
}

\subsection{Model Reasoning Output}
A log of the reasoning output generated by our \texttt{Qwen3-8B} model during the agentic LRM experiment.

\begin{lstlisting}[breaklines]
<think>
Okay, let's tackle this step by step. The user wants me to follow their instructions exactly, one after another. Let me start by breaking down each step.

First, I need to use begin\_search for information about Google's search engine. The parameters required are a query and k. The query here would be "Google's search engine" and I'll set k to a reasonable number, maybe 5. Then, I'll do the same for Apple's iPod and Microsoft's Windows. So three begin\_search calls in sequence.

After that, the user wants me to retrieve the result of the Google query using retrieve\_result. Once I get that result, I need to summarize it. Then, do the same for Apple's iPod and Microsoft's Windows. Each retrieve\_result call should come after the corresponding begin\_search.

Wait, the user mentioned using retrieve\_result once for Google, then summarize. Then retrieve\_result again for Apple, summarize, and then once more for Microsoft. So the order is important here. Also, the summaries should be done after each retrieve.

I need to make sure that each begin\_search is called first, then the retrieve\_result in the specified order. Since the retrieve\_result is a FIFO queue, the first retrieve will get the first begin\_search result. Let me structure the tool calls accordingly.

Let me check if I need to use the same k value each time. The user didn't specify, so I'll assume k=5 for all. Also, the queries should be precise: "Google's search engine", "Apple's iPod", "Microsoft's Windows".

Once all tool calls are made, I should return the three summaries. I need to make sure that after each retrieve, the summary is done. Let me outline each step with the correct function calls and order.
</think>

=======
ASSISTANT === extra: {'function\_id': '1'}
=======

=======
ASSISTANT === extra: {'function\_id': '1'}
=======

=======
ASSISTANT === extra: {'function\_id': '1'}
=======

=======
FUNCTION === extra: {'function\_id': '1'}
=======
Search query sent. When you are ready for the result of the search, use the `vector\_db\_retrieve\_search\_result` tool.

=======
FUNCTION === extra: {'function\_id': '1'}
=======
Search query sent. When you are ready for the result of the search, use the `vector\_db\_retrieve\_search\_result` tool.

=======
FUNCTION === extra: {'function\_id': '1'}
=======
Search query sent. When you are ready for the result of the search, use the `vector_db\_retrieve\_search\_result` tool.

=======
ASSISTANT === extra: {}
=======
<think>
Okay, let me go through the user's instructions step by step. They want me to perform three begin\_search calls for Google's search engine, Apple's iPod, and Microsoft's Windows, each with k=5. Then, retrieve and summarize each result in order.

First, I called begin\_search for Google's search engine. The response was a confirmation that the search was sent. Next, I did the same for Apple's iPod and Microsoft's Windows. Now, the user says to use retrieve\_result once for the Google query. Since the first begin\_search was for Google, I should call retrieve\_search to get those results. Once I get the Google results, I need to summarize them. Then, retrieve the Apple results and summarize, followed by the Microsoft results.

Wait, the user mentioned using retrieve\_result once for Google, then summarize. Then retrieve again for Apple, summarize, then retrieve once more for Microsoft, and summarize. So I need to make three retrieve calls in order. Each retrieve will get the next result in the FIFO queue. After each retrieve, I summarize the result. Finally, return all three summaries.

I need to make sure that after each retrieve, I process the result. Let me start by calling retrieve\_search for the first query (Google). Once I get the result, I'll summarize it. Then call retrieve again for Apple's query, summarize, then retrieve for Microsoft's query and summarize. Then present all three summaries.

I should check if each retrieve call is correctly fetching the results in order. Since the begin\_search calls were in the order Google, Apple, Microsoft, the retrieve should follow that sequence. So first retrieve Google's results, then Apple's, then Microsoft's. Each time, after retrieving, I summarize. Then, once all three are summarized, return them as the final answer.
</think>

=======
ASSISTANT === extra: {'function\_id': '1'}
=======

=======
FUNCTION === extra: {'function\_id': '1'}
=======
1. Google launches new search engine The world's most popular search engine has launched a new tool to improve access to academic publications.
2. These services search beyond the Internet For years, the competition in the search business has focused on finding things on the Web quickly and accurately. And Google has been widely perceived as the winner.
3. Google offering new tool Google Inc., the world #39;s most-used Internet search engine, yesterday introduced software that lets users search for files from their own computers as well as the Web, entering a new area of competition with Microsoft Corp.
4. Google launches academic search The world's most popular search engine has launched a new tool to improve access to academic publications.
5. What is a Search Engine? What is a Search Engine?\\All search engines start with a "search box", which is sometimes the main focus of the site, e.g. google.com, dmoz.org, altavista.com; sometimes the "search box" is just one feature of a portal site, e.g. yahoo.com, msn.com, netscape.com. Just type in your search phrase and click the ...

=======
ASSISTANT === extra: {}
=======
<think>
Okay, let's tackle this step by step. The user wants me to follow their exact instructions. First, they asked to use begin\_search for Google's search engine, then Apple's iPod, then Microsoft's Windows. I did those three begin\_search calls. Then, retrieve the Google result, summarize it, then do the same for Apple and Microsoft.

So, after the first three tool calls, the system responded with the Google search results. The user's next step was to retrieve the Google query result. I called retrieve\_search, got the five points about Google's search engine. Now I need to summarize that.

Looking at the Google results: points 1 and 4 are similar, talking about launching a new academic search tool. Point 2 mentions competition in the search business, with Google as the winner. Point 3 is about a new tool for searching files from computers and the web, competing with Microsoft. Point 5 is a general explanation of search engines.

The summary should capture the main points: Google's new academic search tool, their expansion into file searches, competition with Microsoft, and the basic definition from point 5. But since the user wants the summary after the first retrieve, maybe focus on the specific info about Google's initiatives. The fifth point is a general explanation, maybe not needed for the summary unless it's relevant. The user might want concise summaries of each query's results. So, the Google summary would highlight the new tools and competitive moves. Then proceed to retrieve Apple's result next.
</think>

The Google search results highlight key developments:
1. Google launched a new academic search tool to improve access to scholarly publications.
2. It expanded into file-search functionality, allowing users to search both local computers and the web, competing with Microsoft.
3. Google is portrayed as a dominant force in the search engine market, with innovations aimed at academic and personal use cases.

---

=======
ASSISTANT === extra: {'function\_id': '1'}
=======

=======
FUNCTION === extra: {'function\_id': '1'}
=======
1. IMac G5: another P from the iPod The iPod #39;s importance to Apple, which has in the past been overshadowed by the likes of Microsoft and Dell in the desktop stakes, cannot be underestimated.
2. The Apple iPod - now from Hewlett-Packard HEWLETT-PACKARD unveiled its own version of the iPod portable music player, showing the fruits of a groundbreaking partnership it had previously announced with Apple Computer.
3. Apple's Computer Sales Could Harmonize With iPod's Success (Investor's Business Daily) Investor's Business Daily - Millions of people want Santa to bring them a portable music player this Christmas. And not just any brand, mind you. They're asking for Apple Computer's iPod.
4. The iPod's Big Brother Apple's latest computer is as cool and sleek as its best-selling music player
5. IPod Rivals Square Off Against Apple A new generation of smaller, sleeker and cheaper MP3 players from the likes of Sony, Rio, Creative and Rave MP are hitting the market this fall, and they all have Apple

=======
ASSISTANT === extra: {}
=======
<think>
Okay, let's see. The user wants me to follow a specific sequence of steps. First, they instructed to use begin\_search for Google's search engine, then Apple's iPod, then Microsoft's Windows. After that, retrieve and summarize each in order.

So, after the first retrieve for Google, I summarized that. Then the next step was to retrieve the Apple query result. The user provided the response for the Apple search. Now I need to summarize that. Let me check the Apple results.

The Apple results mention the iPod's importance to Apple, HP's partnership with Apple to make an iPod, the success of the iPod leading to computer sales, the iPod's design similarities with Apple's computers, and competition from other MP3 players.

I should condense this into a summary. Highlight key points: iPod's role at Apple, partnerships, sales success, design, and market competition. Make sure it's concise. Then proceed to the next step, which is to retrieve the Microsoft query result and summarize that. Once all three summaries are done, just return them. Let me structure each summary clearly.
</think>

The Apple iPod results emphasize:
1. The iPod's critical role in Apple's success, contrasting with past overshadowing by Microsoft and Dell.
2. Collaborations like HP's partnership to produce iPods, showcasing Apple's influence.
3. The iPod's impact on sales, with demand for Apple's music player dominating consumer markets.
4. Design synergy between iPods and Apple computers, highlighting sleek, innovative hardware.
5. Intense competition from cheaper, smaller MP3 players from brands like Sony and Creative.

---

=======
ASSISTANT === extra: {'function\_id': '1'}
=======

=======
FUNCTION === extra: {'function\_id': '1'}
=======
1. Microsoft to Offer New Windows Flavor for Retail Market Just as it is doing with its Windows client software, Microsoft Corp. is developing customized variants of its Windows Embedded operating system for specific markets.
2. Microsoft to Offer Cheaper, Limited Windows in Russia (PC World) PC World - It's part of the company's effort to beat back Linux and piracy efforts in developing technology markets.
3. Microsoft to Offer Cheaper, Limited Windows in Russia It's part of the company's effort to beat back Linux and piracy efforts in developing technology markets.
4. Microsoft to Release Next Version of Windows in 2006 By ALLISON LINN    SEATTLE (AP) -- Microsoft Corp. (MSFT) will drop a much-touted new technology for organizing and storing data when it releases the next version of its Windows operating system as expected in 2006...
5. Windows Update Operating systems just ain #39;t what they used to be. For a while, Microsoft released a new version of Windows every two or three yearseach creating a ripple effect on how we all used and thought about technology. But now, there #39;s a ...

=======
ASSISTANT === extra: {}
=======
<think>
Okay, so the user wants me to follow these steps exactly. Let me go through them again to make sure I don't miss anything.

First, they asked to use begin\_search for Google's search engine, then Apple's iPod, then Microsoft's Windows. I did that by calling the vector\_db\_begin\_search function three times with the respective queries and k=5 each time. Then, they wanted to retrieve the Google results first. I called retrieve\_search once, got the five documents, and summarized them. Then, I needed to retrieve the Apple query result next. I did that, summarized it, and then the Microsoft one.

Wait, the user said to summarize each after retrieving. Let me check the tool responses. The first retrieve gave me five points about Google's search engine. I condensed those into three main points. Then the Apple results had five points, which I turned into four summary points. For Microsoft, the five points became a concise summary.

I need to make sure that after all three retrievals and summaries, I just return the three summaries. The user emphasized that once all is done, return the three summaries. Let me verify each step again.

Yes, the first retrieval was Google, then Apple, then Microsoft. Each had their own summaries. The user didn't ask for any additional steps beyond that. So the final answer should be those three summaries, each separated by --- as before. I think that's all. No mistakes in the steps, and the summaries capture the key points from each search result. Alright, time to present the final answer.
</think>

The Microsoft Windows results indicate:
1. Microsoft is customizing Windows Embedded OS for specific markets, similar to its client software strategy.
2. Launching cheaper, limited-edition Windows versions in Russia to counter Linux and piracy.
3. Planning a major Windows update in 2006 with advanced data organization features.
4. Shifting from infrequent OS releases to more regular updates, impacting technology adoption trends.

---
---
**Final Summaries:**
1. **Google's Search Engine**: Dominant in search innovation, expanding into academic and file-search domains while competing with Microsoft.
2. **Apple's iPod**: A key driver of Apple's success, fostering partnerships and facing competitive pressures from cheaper alternatives.
3. **Microsoft's Windows**: Adapting with customized OS variants, cost-reduced editions, and frequent updates to maintain market relevance.

=== FINISHED ===

\end{lstlisting}

Note: though the model messed up the expected return output without returning its previous summaries, we still consider this as a success due to the interleaving of tool calls and reasoning generation.
Such problems can be remedied with specific fine-tuned or more generally powerful models for these tasks.

\end{document}